\documentclass[twocolumn]{revtex4}

\usepackage{setspace} 

\usepackage{amsmath}

\newcommand{\bs}{{b_{\rm s}}}   
\newcommand{\bd}{{b_{\rm d}}}   
\newcommand{\tbs}{{\bar{b}_{\rm s}}}   
\newcommand{\tbd}{{\bar{b}_{\rm d}}}   
\newcommand{\eps}{{\varepsilon_{\rm b}}}   
\newcommand{\Nbp}{{n}}          
\newcommand{\Eini}{{\varepsilon_{\ell}}}  
\newcommand{\cS}{{\cal{S}}}        
\newcommand{\fc}{{f_c}}           
\newcommand{\tfc}{{\tilde{f}_c}}  
\newcommand{\fd}{{f^*}}           
\newcommand{\tw}{\tau_{\rm w}}    
\newcommand{\ts}{\tau_{\rm s}}    
\newcommand{\tc}{\tau_{\rm c}}    
\newcommand{\ttc}{\tilde{\tau}_{\rm c}}    
\newcommand{\kin}{k_{\rm in}}     
\newcommand{\kout}{k_{\rm out}}     

\usepackage{graphicx}

\usepackage{natbib} 
\begin{document}
\title{ DNA as a programmable viscoelastic nanoelement}

\author{Richard A. Neher} 
\email{Richard.Neher@physik.lmu.de}

\author{Ulrich Gerland}
\email{Ulrich.Gerland@physik.lmu.de}

\affiliation{Arnold Sommerfeld Center for Theoretical Physics (ASC) and Center for Nanoscience (CENS), Ludwig-Maximilians Universit\"at M\"unchen, Theresienstrasse 
37, 80333 M\"unchen, Germany}

\date{\today}

\begin{abstract} 
The two strands of a DNA molecule with a repetitive sequence can pair into many different basepairing patterns. For perfectly periodic sequences, early bulk experiments of P\"orschke indicate the existence of a sliding process, permitting the rapid transition between different relative strand positions [{\it Biophys. Chem.} {\it 2} (1974) 83]. Here, we use a detailed theoretical model to study the basepairing dynamics of periodic and nearly periodic DNA. As suggested by P\"orschke, DNA sliding is mediated by basepairing defects (bulge loops), which can diffuse along the DNA. Moreover, a shear force $f$ on opposite ends of the two strands yields a characteristic dynamic response: An outward average sliding velocity $v\sim 1/N$ is induced in a double strand of length $N$, provided $f$ is larger than a threshold $\fc$. Conversely, if the strands are initially misaligned, they realign even against an external force $f<\fc$. These dynamics effectively result in a viscoelastic behavior of DNA under shear forces, with properties that are programmable through the choice of the DNA sequence. We find that a small number of mutations in periodic sequences does not prevent DNA sliding, but introduces a time delay in the dynamic response. We clarify the mechanism for the time delay and describe it quantitatively within a phenomenological model. Based on our findings, we suggest new dynamical roles for DNA in artificial nanoscale devices. The basepairing dynamics described here is also relevant for the extension of repetitive sequences inside genomic DNA. 
\end{abstract}


\maketitle




\section*{Introduction}
The basic double-helical structure of DNA is insensitive to the 
nucleotide sequence, but many of its biophysical properties are not. 
For instance, the local thermodynamic stability of double stranded DNA 
(dsDNA) depends strongly on the sequence \citep{SantaLucia:98}, and certain sequence motifs can cause permanent bends or make DNA more bendable \citep{Hagerman:90}. Such local modulations of the DNA properties play an important role in molecular biology, e.g. for nucleosome positioning \citep{Widom:01} and transcription regulation through DNA looping \citep{Rippe:95}. The sequence-dependent stability of DNA basepairing is also crucial for applications in nanotechnology \citep{Seeman:03,Albrecht:03,Lin:04}. Clearly, since the thermodynamics of DNA basepairing is sequence-dependent, the kinetics is sequence-dependent as well. Our aim here is to show that the kinetics can display a much richer phenomenology than might be expected on the basis of the thermodynamics alone. 

\begin{figure}[b]
\includegraphics[width=8cm]{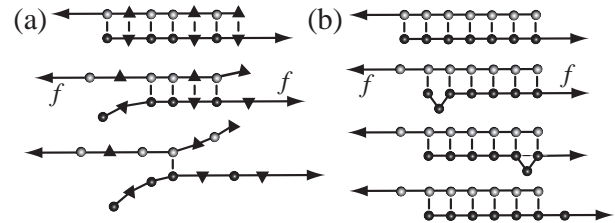}
\caption{\label{FIGshearing}DNA under a shear force. 
(a) A non-periodic sequence unravels from both ends, driven by the length gain from converting stacked bases into longer single strand. 
(b) A periodic DNA sequence can open by sliding, mediated by bulge loops that are created at the ends and diffuse freely along the DNA. When a bulge loop reaches the opposite end, the two strands have effectively slipped against each other by a distance equal to the loop size. 
}
\end{figure}

The dynamics of DNA basepairing can be probed experimentally on the 
single-molecule level with mechanical and optical techniques 
\citep{Bustamante:00,Clausen-Schaumann:00,Merkel:01,Lavery:02,Bockelmann:02,Danilowicz:03,Altan-Bonnet:03}. One approach is to unzip dsDNA from one end of the double helix \citep{Lubensky:02,Bockelmann:02,Danilowicz:03}. However, unzipping probes only one aspect of the basepairing dynamics, the sequential opening of consecutive basepairs. In a different approach, a shear force is applied by grabbing the two strands on opposite ends of the dsDNA \citep{Strunz:99}, see Fig.~\ref{FIGshearing}. For a heterogeneous dsDNA with a random sequence, the effect of a shear force is to unravel the basepairs from both ends \citep{Strunz:99}, see 
Fig.~\ref{FIGshearing}a, which is qualitatively similar to unzipping. 
In contrast, with a perfectly periodic sequence, e.g. $(C)_N$ on the upper and $(G)_N$ on the lower strand or a higher order repeat such as $(CA)_N$ and $(GT)_N$, the two strands can bind in many configurations \citep{Poerschke:74}. An applied shear force then facilitates local strand slippage and can induce macroscopic DNA sliding \citep{Neher:04}, see Fig.~\ref{FIGshearing}b. [Throughout this paper, we use the term `DNA slippage' for microscopic events where a few bases at the end of the double strand unbind and rebind shifted by one or several repeat units. In contrast, `DNA sliding' refers to an average large-scale movement of the two strands against each other.]

DNA slippage is an aspect of DNA basepairing dynamics, which plays an 
important role in the generation of a class of genetic diseases 
\citep{Reddy:97,Lovett:04}. If local DNA slippage occurs in an Okazaki fragment during DNA replication, trinucleotide repeats inside genes can get extended beyond a threshold length for the onset of Huntington's and other diseases. Such slippage events are possible only within the time window that DNA polymerase needs to fill in the Okazaki fragment. Thus, the kinetics of strand slippage is an important determinant for the frequency of repeat extensions. 

We propose that the dynamics of periodic and nearly periodic DNA is interesting also for the design of DNA-based nanodevices. Indeed, DNA is becoming increasingly popular as a building block for the assembly of nanoscale structures and devices \citep{Seeman:03,Albrecht:03,Lin:04}. These applications already exploit the specificity of the basepairing interaction, e.g. to direct the assembly of DNA strands into predefined architectures, and the dynamics of DNA branch migration, e.g. to replace a bound DNA strand by a different strand. In the discussion section, we consider several new possible applications of DNA in nanotechnology, based on the dynamic properties identified in the main part of this paper. 

Finally, DNA sliding is interesting also from a purely theoretical perspective. Since simultaneous slippage of all basepairs is kinetically inhibited by an extensive activation barrier, the macroscopic sliding of DNA strands is a complex process involving the dynamics of many local basepairing defects \citep{Poerschke:74,Neher:04}. The most likely defects are bulge loops, see Fig.~\ref{FIGshearing}b, which are created at the ends of the dsDNA, or, in pairs, anywhere along the molecule. Once formed, bulge loops diffuse freely along a periodic dsDNA until they annihilate with a loop on the opposite strand or are absorbed at an end. Mutations in the periodicity of the DNA sequence create obstacles for the diffusion of bulge loops. Effectively, the bulge loop dynamics can be regarded as a reaction-diffusion process of particles and antiparticles in one dimension. DNA shearing experiments render certain aspects of these dynamics observable and permit a quantitative characterization. 

The outline of this paper is as follows. First, we describe our theoretical model for the energetics and dynamics of DNA basepairing under a shear force. We then show that our model leads to the following predictions: 
(i) For periodic dsDNA, the combination of polymer mechanics and 
basepairing dynamics gives rise to a viscoelastic response to shear 
forces above a threshold $\fc$, where both $\fc$ and the viscosity $\eta$ are programmable over a wide range through the DNA sequence. The viscoelastic behavior can be described with the help of a mechanical analog model. 
(ii) DNA sliding is possible even when the exact sequence periodicity is destroyed by a few mutations. 
(iii) The mutations affect the viscoelastic behavior by introducing a programmable time delay before sliding commences after a sudden force jump.  
(iv) The mechanism for the time delay can be understood within a 
phenomenological model, which also permits a quantitative description of the full distribution of time delays. Taken together, we find that the sequence dependence of the basepairing dynamics allows to adjust the mechanical response of DNA under a shear force over a broad range of behaviors. In the last section, we discuss the experimental ramifications of these findings.

\section*{DNA model} 

To study the dynamics of DNA sliding, we consider a DNA molecule under a shear force $f$, which can be applied experimentally by pulling the opposite 5' ends \citep{Strunz:99} or, alternatively, the opposite 3' ends. In a coarse-grained description, the configuration of the DNA is specified by its basepairing pattern $\cS$ and the spatial contours of both strands. A generic configuration consists of two unstretched and two stretched single strands, and the central region from the first to the last basepair, see Fig.~\ref{FIGshearing}. 

We will not explicitly describe the dynamics of the spatial polymer degrees of freedom, but assume rapid equilibration compared to the timescale of DNA sliding. This assumption is justified for short DNA molecules: The timescale to equilibrate a semiflexible polymer of length $L$ and persistence length $l_p$ in a solvent of viscosity $\eta$ is $\eta L^4/72\, {l_p}^2f$, where $f$ is an external force applied to its ends \citep{Hallatschek:05}. For a DNA of 150 bp (one persistence length) in water at a 10~pN load, the equilibration time is on the order of $0.01\,\mu$s, which is fast compared to the millisecond timescale of DNA sliding observed in the reannealing experiments of P\"orschke \citep{Poerschke:74}. Hence, we integrate out the contour conformations to obtain a free energy function $E(\cS)$ that depends only on the basepairing pattern $\cS$. The total free energy $E(\cS)$ can be split up into three terms, 
\begin{equation}
\label{FreeEnergy}
E(\cS)=E_{\mathrm stretch}(\cS)+E_{\mathrm bp}(\cS)+E_{\mathrm loop}(\cS) \;,
\end{equation}
corresponding to the stretching energy, the free energy gain due to basepairing, and the free energy cost of (internal or bulge) loops in the pattern $\cS$, respectively. 

\begin{figure}[tb]
\includegraphics[width=8cm]{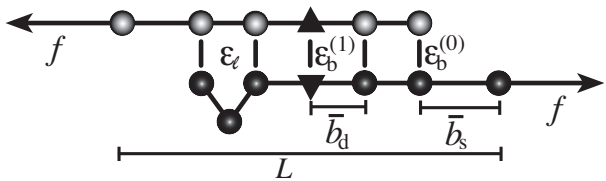}
\caption{\label{FIGmodel} DNA free energy model. The free energy $E(\cS)$ of a basepairing pattern $\cS$ contains three separate contributions: (i) A negative binding energy for basepairing. For simplicity we assign the same binding energy $\eps^{(k)}$ for every basepair of type $k$, regardless of the neighboring bases. (ii) A positive free energy cost for internal and bulge loops. We assign the same cost $\Eini$ for every loop, regardless of its length and base sequence, since the detailed choice of the loop cost function does not affect our main findings. (iii) A stretching energy. For a given pattern $\cS$, the stretching energy can be written in the form $- f\,L(\cS)$, with an effective length $L(\cS)$, which is obtained from force-dependent base-to-base distances $\tbd(f)$ and $\tbs(f)$ for double and single strand, respectively. Note that $L(\cS)$ does not correspond to the physical length of the DNA molecule (see main text).
}
\end{figure}

{\it Polymer model.---}
The mechanical polymer properties of DNA enter only into the stretching energy, which we write in the form $E_{\mathrm stretch}(\cS)=-f\,L(\cS)$, with an effective force-dependent total length $L(\cS)$ for the stretched DNA, i.e. the two single-stranded ends where the force is applied and the central DNA segment from the first to the last base pair, see Fig.~\ref{FIGmodel}. The unstretched single strands do not contribute to the free energy, since we take all energies relative to the unstretched and unpaired state, which is the usual convention \citep{SantaLucia:98}. For the stretched single strands, we use a freely jointed chain polymer model with a Kuhn length twice the bare segment length $\bs\approx0.7$~nm for a single base \citep{Smith:96}. With this model, each unbound base at the ends where the force is applied contributes an effective length $\tbs(f)$ to the total length $L(\cS)$, where 
\begin{equation}
\tbs(f)=-\frac{k_{B}T}{2f}\ln\left(\frac{\sinh(2f\bs/k_{B}T)}{2f\bs/k_{B}T}\right)
\end{equation}
and $k_{B}T$ is the thermal energy. Note that $\tbs(f)$ differs from the average extension of one segment in the direction of the force. Instead, the average total extension $\langle x\rangle$ of a DNA with basepairing pattern $\cS$ is calculated as the force derivative of the stretching free energy, 
\begin{equation}
\langle x\rangle = \frac{\partial E_{\mathrm stretch}(\cS)}{\partial f}\;,
\end{equation}
which yields the correct (Langevin) form for the extension of a freely jointed chain. For the central DNA region from the first to the last basepair, we assume a B-DNA conformation and use a wormlike chain model with persistence length $l_{p}=50$~nm and a contour length of $\bd=0.34$~nm per base. [The length of an asymmetric loop in the central region is approximated by counting only the bases in the shorter arm of the loop.]
For the forces of interest here, the effective length of a basepair, $\tbd(f)$, is given by the asymptotic formula
\begin{equation}
\tbd(f)=\bd\left(1-\sqrt{\frac{k_{B}T}{4f l_{p}}}\right)\;.
\end{equation}
The force-dependence of the lengths $\tbs(f)$, $\tbd(f)$ is in fact essential only for our calculation of the viscoelastic response. For all other properties considered below, the force-dependence has no qualitative effect, and will hence be neglected (i.e. $\tbs(f)=\bs$, $\tbd(f)=\bd$ everywhere except in the section `Viscoelastic behavior'). 

{\it Base-pairing energy model.---}
To obtain a compact theoretical description, we use a base-pairing energy model which is simplified from the nearest-neighbor model of Ref.~\citep{SantaLucia:98}, but nevertheless permits semi-quantitative predictions. We exclude base pairs within a strand, and assign a binding (free) energy $\eps^{(k)}>0$ for each base pair of type $k$ (Watson-Crick or other) between strands, see Fig.~\ref{FIGmodel}. Hence, $E_{\mathrm bp}(\cS)=-\sum_{k}\Nbp_{k}(\cS)\,\eps^{(k)}$, where $\Nbp_{k}(\cS)$ is the total number of type $k$ basepairs in the base-pairing pattern $\cS$. Similarly, we assign a loop initiation cost $\Eini>0$ for each internal or bulge loop in a given pattern (we neglect an additional length-dependent loop cost, which has no qualitative effect on the results discussed below). Therefore, $E_{\mathrm loop}(\cS)=q(\cS)\,\Eini$, where $q(\cS)$ is the total number of loops in the pattern. The numerical values of the free energy parameters $\eps^{(k)}$ and $\Eini$ are temperature dependent \citep{SantaLucia:98}. The actual values used in our simulations are given below. 

{\it Elementary kinetic steps.---}
We support our phenomenological theory presented below by simulating the DNA base-pairing dynamics in detail. To this end, we use a kinetic Monte Carlo scheme with three single base moves as elementary steps \citep{Flamm:00}: basepair opening, basepair closing, and `basepair slippage'. Here, basepair slippage refers to a local shift of the binding partner of a base, which is possible only for basepairs next to unbound bases, i.e. inside loops or at the ends of the molecule. Clearly, basepair slippage can also be generated by a basepair opening move followed by a basepair closing move. However, the work of P\"orschke \citep{Poerschke:74} indicates that basepair slippage is faster than would be expected from the individual rates for basepair opening and closing, see below. Hence, we include the basepair slippage move, as has been done previously \citep{Flamm:00}. 

{\it Kinetic rates.---}
To fully specify our model for the DNA base-pairing dynamics, we need to assign a rate to each elementary kinetic step. Careful relaxation experiments \cite{Craig:71} determined the rate for basepair closing at the end of helical segments to be $1-20\times 10^6\,s^{-1}$, where the range indicates the experimental uncertainty. In our model, we assume that this rate is independent of the identity of the basepair. To reproduce the correct equilibrium behavior from our base-pairing dynamics, the rate for opening a basepair of type $k$ at the end of a helix must be reduced by a factor $\exp(-\eps^{(k)}/k_{B}T)$ with respect to the closing rate. From reannealing experiments with periodic sequences, P\"orschke \cite{Poerschke:74} estimated the rate for the displacement of a bulge loop by one base, i.e. the rate for basepair slippage, to be roughly $5\times 10^6\,s^{-1}$. Hence, the rates for basepair closing and basepair slippage are roughly equal, within experimental accuracy. In our model, we set them exactly equal, for simplicity (our main results are in fact insensitive to the precise value of the closing rate, see below). In general terms, our model assumes that all kinetic rates of passing from a higher energy configuration to one with lower energy are the same, while the reverse rates are chosen to obey detailed balance. It may be noteworthy that recent theoretical work on the kinetics of force-induced RNA unfolding, which used similar assumptions, produced surprisingly good agreement with experiment \citep{Harlepp:03}. 

Below, we report all of our kinetic simulation data in units of Monte Carlo steps (MCS). From Ref.~\citep{Poerschke:74}, our best estimate for the real time equivalent of one Monte Carlo step is $0.2\,\mu$s. However, it should be kept in mind that this estimate bares a large uncertainty.

\section*{Sliding dynamics of periodic sequences}

The simplest periodic sequence is a repetition of one base on one strand, 
e.g. {\tt AAA...},  and the complementary base on the other. In this case, we have only basepairs of one type (i.e. $\eps^{(k)}\equiv\eps$ in our model) and each base on one strand can bind to any base on the other strand. For longer repeat units, e.g. triplet repeats such as {\tt CAGCAG...}, which play an important role in genetic diseases \citep{Reddy:97}, one can treat a repeat unit as an effective base with larger associated energies $\eps$, $\Eini$ and lengths $\bd$, $\bs$. We are interested in the basepairing dynamics induced by a constant shear force $f$ that is suddenly turned on at $t=0$. In the following, we first review the physical description of DNA sliding dynamics, which we have established already in Ref.~\citep{Neher:04}. We then construct a mechanical analog model to characterize the viscoelastic response of periodic DNA and its sequence-dependence. 

{\it Quantitative phenomenological description.---}
As illustrated in Fig.~\ref{FIGshearing}b, sliding of periodic dsDNA 
is mediated by the creation, diffusion, and annihilation of bulge loops. When a force is applied, the diffusion of bulge loops within the dsDNA remains unbiased, assuming the force does not deform the dsDNA structure significantly (this assumption clearly breaks down for forces above the B-S transition around 65~pN). In contrast, the force strongly affects the rates at which bulge loops are created at the ends. When the two DNA strands are misaligned, these creation rates are imbalanced, since a bulge at an overhanging end does not reduce the number of basepairs in the structure (whereas it does on the opposite end). This imbalance produces a restoring force $\fc$, which can be obtained approximately \citep{Neher:04} by balancing the energy cost of breaking a basepair with the gain in stretching energy, 
$\eps=f\cdot(2\,\tbs(f)-\tbd(f))$. The restoring force creates an 
average inward drift that realigns the two strands. To obtain an outward drift velocity $v$, i.e. macroscopic sliding, one needs to overcome the restoring force $\fc$, so that $v \sim (f-\fc)$ in the vicinity of $\fc$. Indeed, $\fc$ becomes a critical force in the thermodynamic sense when the limit of a large strand length $N$ is taken and the state where the strands are completely separated is excluded. 


At the critical force the rates at which bulge loops are produced, are equally large on the overhanging stretched and the unstretched ends. The average sliding velocity $v$ vanishes, however the bulge loop dynamics still leads to a macroscopic diffusion of the two strands relative to each other, with a diffusion coefficient $D$. Interestingly, this diffusion coefficient scales with the total number of bases as $D \sim 1/N$, so that the rupture time $\tau$ required to separate the two strands completely scales as $\tau\sim N^3$ instead of the usual $\tau\sim N^2$ for diffusion of a particle over a distance $N$. This scaling of $D$ is due to the fact that loops are generated at the ends with a constant rate, but only result in a global shift between the strands, if they diffuse over a distance $\sim N$, either to annihilate at the other end or with a loop on the opposite strand. In both cases, the probability for an event scales as $1/N$. The $D \sim 1/N$ scaling occurs also in the reptation problem of polymer physics, and indeed the microscopic origin is closely related, as motion is mediated by defect diffusion \citep{deGennes:71}. 

Since the production of a loop on the stretched ends shortens the molecule, the corresponding production rate decreases with $f$. Hence, the rates of events extending or shortening the double stranded region, that are equal for $f=\fc$, differ at other forces resulting in a drift. Each of these rates, and consequently also the sliding velocity $v$ is proportional to $1/N$. From the Einstein relation one expects $v\sim (f-\fc)D\sim 1/N$, in agreement with this result. 
With the negative (inward) drift velocity below $\fc$, rupture events 
are driven by rare fluctuations, and the rupture time $\tau$ grows 
exponentially with $N$, as is characteristic for thermally activated 
transport over an extensive energy barrier. 
On the other hand, for forces larger than $\fc$, the $N^{-1}$ scaling 
of $v$ leads to rupture times increasing as $\tau\sim N^2$. 

This scaling holds up to a force $\fd$, above which the rupture 
times grow only linearly with $N$, 
due to a dynamical transition in the opening mode from sliding to 
unraveling (i.e. the opening mode of heterogeneous dsDNA). For $f>\fd$, it is energetically favorable to break basepairs consecutively from both ends and both strands are after breaking $N$ basepairs.
Within our model, $\fd$ is well approximated by the solution of 
$f=\eps/[\tbs(f)-\tbd(f)]$, i.e. the balance between the basepairing energy and the stretching energy gained by extending the molecule by the difference between the length of an unbound base and a basepair. [For DNA sequences where this force is large enough to deform the DNA structure, in particular for $\fd$ above the $\sim 65$~pN of the B-S transition, the unraveling mode may not exist.]

{\it Viscoelastic behavior.---}
DNA sliding can be regarded as a viscous flow of the two strands relative to each other. According to the physical picture reviewed above, this flow has interesting nonlinear and sequence-dependent properties. Since the shear force elicits also an elastic response (due to the entropic elasticity of DNA), the behavior of periodic dsDNA is reminiscent of a viscoelastic material. Such materials combine solid-like and fluid-like mechanical properties when probed by external stress. In the following we examine this analogy more closely. 

\begin{figure}[tb]
\includegraphics[width=8cm]{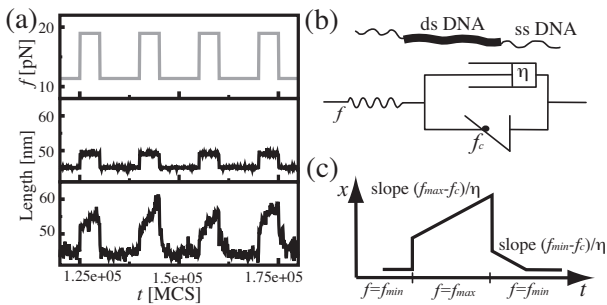}
\caption{\label{FIGviscoelasticity} Viscoelastic response of periodic DNA. 
(a) The shear force on an 80 bp dsDNA (with two 20 bp ssDNA linkers) is switched periodically between $f_{min}=11.4$~pN and $f_{max}=19$~pN (upper panel). The center and bottom panels show the extension-time trace for heterogenous and periodic DNA, respectively (energy parameters: $\eps=1.11\,k_BT$ and $\Eini=2.8\,k_BT$, roughly corresponding to AT basepairs at $50 ^\circ\mathrm{C}$ \citep{Neher:04}). 
The time units are Monte Carlo steps (MCS), the real time equivalent of which is discussed in section `Kinetic rates'.
The heterogeneous DNA responds only elastically to the force jumps, mostly due to stretching of the linkers. The length of the periodic DNA shows a similar elastic strain, but in addition, the molecule elongates at a finite speed due to sliding, since $f_{max}>\fc=16.3$~pN. When the molecule is relaxed, we find an elastic response and inward sliding, since $f_{min}<\fc$.
The length of the periodic DNA fluctuates strongly due to loop formation and annihilation.
(b) The viscoelastic behavior displayed by a periodic DNA molecule can be 
described by a generalized Zener model, where harmonic springs are
substituted by anharmonic elastic elements describing polymer elasticity 
and the restoring force $\fc$. The ideal dashpot (with viscosity $\eta$) creates the viscous behavior of periodic dsDNA. 
(c) The response of the above idealized model to the same periodic force  
resembles the average response of periodic DNA. 
}
\end{figure}

The mechanical behavior of a typical viscoelastic material can be described by a Zener model \citep{Zener:56}, which is constructed e.g. by connecting a Kelvin element (a dashpot in parallel with a spring) in series with a spring, see Fig.~\ref{FIGviscoelasticity}. The Zener model reproduces the two prominent characteristics of viscoelastic 
materials: 
(i) In a `creep experiment', where a constant stress is suddenly applied, an instantaneous elastic strain is followed by a gradual creep towards a new equilibrium. 
(ii) When the strain is suddenly increased, the stress rises sharply and then relaxes gradually to an equilibrium value. 
On a qualitative level, periodic dsDNA displays these same characteristics in its average behavior: 
(i) Upon sudden application of a constant force $f$ in the range 
$\fc<f<\fd$, the DNA rapidly stretches against its entropic elasticity 
and slowly creeps with a viscosity $\eta$ that is proportional to 
the number of bases in the double strand. However, it will not approach a new equilibrium but rupture eventually.  
(ii) When the extension of the DNA is suddenly increased, the tension rapidly rises and then slowly relaxes to the critical value $\fc$ (provided the initial rise was above $\fc$). 

The viscoelastic behavior of periodic DNA can be described by a nonlinear generalization of the Zener model, see Fig.~\ref{FIGviscoelasticity}b, where the Kelvin element effectively describes the basepairing dynamics, while the outer elastic element accounts for the entropic elasticity of the polymer backbone (consisting of dsDNA, ssDNA, and, if present, linkers to the points of force application). Since the basepairing dynamics of two misaligned complementary periodic DNA strands produces a restoring force $\fc$, the sliding velocity $v$ is proportional to $f-\fc$. The sliding dynamics is thus described by a dashpot in parallel with a potential generating the restoring force and preventing contraction beyond maximal overlap. In contrast to the standard Zener model with harmonic springs, the stress in response to a strain, will relax to the value $\fc$, independent of the displacement (within a certain range). 
Fig.~\ref{FIGviscoelasticity}a shows extension-time traces obtained from our model, both for a periodic (bottom panel) and a heterogeneous DNA (center panel), see caption for parameters. Here, we have considered a creep test situation where the force is switched periodically between $f_{min}<\fc$ and $f_{max}>\fc$ (top panel). 
Fig.~\ref{FIGviscoelasticity}c shows the corresponding behavior of 
the generalized Zener model for comparison. 
We observe that the average behavior of the periodic DNA resembles that of the generalized Zener model, while the heterogeneous DNA shows only elastic behavior. Of course the extension also displays strong thermal fluctuations, which play an important role in single molecule dynamics, and ultimately lead to rupture even below the critical force \citep{Neher:04}.

{\it Programmability.---}
The viscoelastic behavior described above relies on the basepairing dynamics within the DNA molecule, and is manifestly sequence-dependent. This fact makes the mechanical behavior of dsDNA under shear force programmable, i.e. both the force offset $\fc$ and the viscosity $\eta$ can be adjusted through sequence composition and length of the dsDNA. 
Even for perfectly periodic sequences, there is still a considerable 
freedom in the choice of the sequence composition, since a repeat unit can be several bases long and involve different combinations of Watson-Crick and other basepairs. 
Exploiting this freedom, the range over which the average basepairing 
energy $\eps$ can be programmed is roughly $0.5-4\,k_BT$ \citep{SantaLucia:98}, which translates into an equally wide range of force 
offsets\footnote{The precise experimental range of the force offset is 
difficult to predict, since it depends sensitively on the effective 
ssDNA and dsDNA length. Roughly, we expect values of up to 30~pN.} 
$\fc=\eps/(2\,\tbs-\tbd)$. 

The velocity of macroscopic strand sliding is determined by four factors: (i) the mobility of defects, i.e. the rate for bulge loop displacement, (ii) the bulge loop density, (iii) the inverse strand length, and (iv) the deviation of the force $f$ from the critical force $\fc$. The defect density depends sensitively on the base-pairing free energy and may vary roughly between 0.001 and 0.2 for different repeat lengths and temperatures, leaving great freedom to adjust the timescale of DNA sliding. Note that since only the bulge loop density and not the individual rates for basepair closing and opening influence the sliding velocity, the rate for bulge loop displacement is the only crucial rate parameter in our model. By increasing the strand length, the sliding velocity can be made arbitrary small, or, equivalently, the viscosity $\eta$ can be made arbitrarily large ($\eta\sim N$). Alternatively, $\eta$ is increased by using longer repeat units, since $\eta$ grows exponentially with the free energy cost of creating a bulge loop. 
An order of magnitude estimate\footnote{The reannealing experiments of 
Ref.~\citep{Poerschke:74} suggest that a misaligned 10~bp molecule can 
slide by one basepair within $0.1$~ms. Assuming that the sliding velocity extends linearly from $\fc\sim 10$~pN to force zero, one obtains the estimate $\eta\sim 3\cdot 10^{-3}$~pN$\cdot$s/nm.} 
for the lower bound on $\eta$ yields $\sim 10^{-3}$~pN$\cdot$s/nm, based on reannealing experiments with homogeneous oligonucleotides of 10~bps \citep{Poerschke:74}. With these force- and timescales, DNA sliding should be well observable in single-molecule experiments.

\section*{Periodic DNA with weak sequence disorder}

\begin{figure}[tb]
\includegraphics[width=8cm]{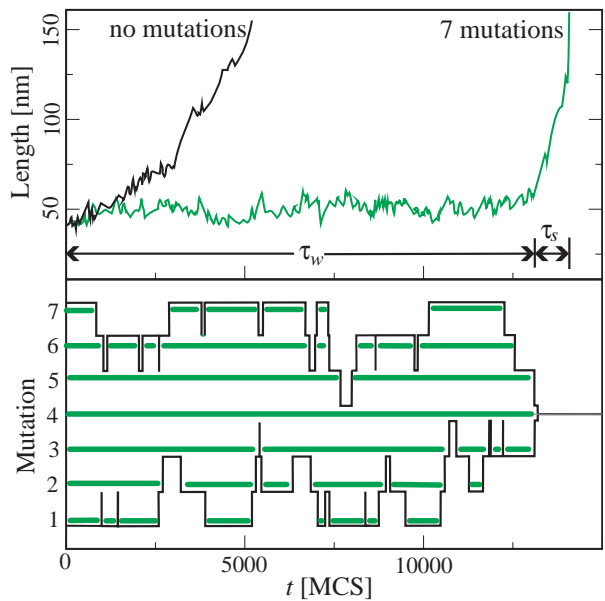}
\caption{Top: 
Extension-time trace for a perfectly periodic DNA of $N=120$ basepairs and the same DNA with $7$ mutations, both under a shear force of $f=12.7$~pN (energy parameters as in Fig.~\ref{FIGviscoelasticity}). Whereas the molecule without mutations starts sliding almost immediately, the molecule with mutations fluctuates about its initial length for some time $\tw$ before sliding starts.
Bottom: 
The time trace of the binding state (open/closed) for the seven mutated basepairs in the sequence. Each mutated basepair (1 to 7) is unbound wherever the corresponding thick horizontal line is broken, and bound where the line is shown. Note that the mutated basepairs do not open/close independently from each other. Instead, a mutated basepair opens only once all mutated basepairs to the left or right are already open. The black envelope curves emphasize the positions of the outermost bound mutation on each side. Their dynamics resembles a (biased) random walk. Sliding begins when all mutations are open. 
}
\label{FIGtrajectories}
\end{figure}

How is the basepairing dynamics affected when a few mutations destroy 
the perfect periodicity? 
Fig.~\ref{FIGtrajectories} (top) shows two simulated extension-time traces, 
one for a perfectly periodic sequence and one with $M=7$ equidistant 
mutations (DNA parameters: see caption). 
Here, we assigned the same binding energy to mutated and original 
basepairs, in order to focus on the effects that mutations exert on the 
basepairing dynamics rather than the energetics. 
Furthermore, we assumed that mutated bases can only bind to their 
ground-state binding partners, i.e. mutated bases cannot form basepairs 
with the original bases and all mutations are of a different type. 
The less generic effects that can result without these assumptions are discussed below. 

We observe that the mutations have a drastic effect: 
whereas the original sequence begins to slide almost immediately 
after application of the force, the mutated sequence exhibits a 
pronounced delay before sliding sets in. 
Indeed, the figure suggests that the mutated sequence has two 
characteristic timescales, a waiting time $\tw$ during which the 
extension fluctuates around a constant value, and a sliding time $\ts$ 
during which the extension increases until the two strands are 
completely separated. 
A second, less drastic effect of the mutations is to reduce $\ts$, i.e. 
once sliding starts, it is faster than without mutations. 
Note that the convex shape of both sliding curves is due 
to the fact that the sliding velocity increases as the length of the 
double stranded region decreases, $v\sim\eta^{-1}\sim N^{-1}$, see above. 

What is the physical mechanism that sets the waiting timescale $\tw$?
Clearly, sliding can begin only after all mutated basepairs have been 
broken, since otherwise the two strands are locked into one relative 
position. 
Arguably the simplest scenario would be that all mutations independently 
fluctuate between the open and closed state, and sliding commences when 
all mutations happen to be open simultaneously. 
Alternatively, the dynamics of the mutated basepairs could be correlated. 
To clarify the dynamical mechanism, we plotted the binding state 
(bound/unbound) of all mutated basepairs alongside the trajectories in 
Fig.~\ref{FIGtrajectories} (bottom, gray curves). 
It is evident that the mutations do not open independently. 
Instead, interior mutations open only once the neighboring mutation 
towards the exterior has already opened.

\begin{figure}[tb]
\includegraphics[width=8cm]{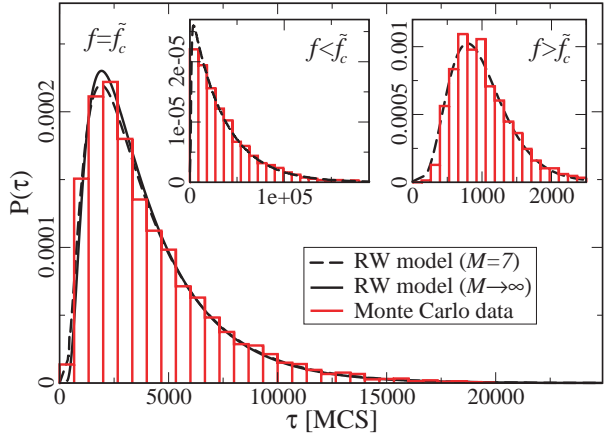}
\caption{Waiting time distributions. Main panel: The histogram of waiting times $\tw$ of a 120 bp long DNA sequence with $M=7$ equidistant mutations subject to a force $f=\tfc=12.9$~pN, is well described by the distribution of collision times (dashed line) of the two random walker (RW) model, see main text and Fig.~\ref{FIGRandomWalks}. The solid line shows the parameter-free asymptotic distribution of Eq.~(\ref{critical_distribution}) for comparison.
Insets: Distribution of $\tw$ for forces above and below $\tfc$ ($f=15.2$~pN and $f=11.4$~pN, respectively). The dashed lines are fits using the RW model with directional bias, see main text. 
}
\label{FIGtwait_distribution}
\end{figure}

{\it Two random walker model.---} 
Inspection of Fig.~\ref{FIGtrajectories} (bottom) suggests that the 
positions of the two outermost bound mutations might in fact perform a 
(biased) random walk, see the black curves in Fig.~\ref{FIGtrajectories}. 
If true, the waiting time $\tw$ could be interpreted as the first 
collision time $\tc$ of two random walkers (2RW) on a row of $M+1$ discrete 
sites, with force-dependent in- and outward hopping rates $\kin$ 
and $\kout$. 
To test this hypothesis, we compare the histogram of $\tw$'s 
(from many simulations) with the distribution $P(\tc)$ of first 
collision times for 2RW. 
Fig.~\ref{FIGtwait_distribution} shows three such histograms (main panel 
and two insets) obtained with the same DNA parameters as in 
Fig.~\ref{FIGviscoelasticity}, but with three different forces. 
Superimposed are the distributions $P(\tc)$, 
calculated as described below and in {\it Supporting Material}.  
The case shown in the main panel of Fig.~\ref{FIGtwait_distribution} 
corresponds to a force value for which the 2RW are unbiased, 
i.e. $\kin=\kout$, whereas the left inset corresponds to a smaller force 
producing a bias to the outside ($\kout>\kin$) and the right inset shows 
the opposite case of a larger force and $\kout<\kin$. 
In all three cases, the observed histogram is well described by the 2RW model. 
Indeed, despite some caveats (see below), this model can serve as a 
useful coarse-grained description for the basepairing dynamics preceding 
the sliding stage. 

\begin{figure}[tb]
\includegraphics[width=8cm]{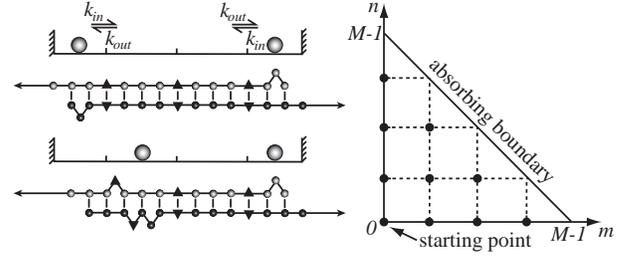}
\caption{On a coarse-grained level, the dynamics of mutation opening/closing can be described by a model of two particles hopping on a 1D lattice, with inward/outward hopping rates $\kin$, $\kout$. Their positions represent the two outermost closed mutations. When the particles collide, all mutations have opened. Equivalently, one can consider a single particle hopping on a triangular 2D lattice. The first collision time then corresponds to the time to reach the diagonal absorbing boundary. 
}
\label{FIGRandomWalks}
\end{figure}

The calculation of the first collision time distribution $P(\tc)$ belongs 
to the class of first passage problems, which has been studied extensively 
in statistical physics \citep{Gardiner:83}. 
In the context of the helix-coil transition, Schwarz and Poland \citep{Schwarz:75} (see also \citep{Anshelevich:84}) already solved the associated diffusion problem. Here, we use their work as a basis to treat the first passage problem. 
One can replace the problem of 2RW in one dimension by the equivalent problem of one RW on a two dimensional lattice with a triangular shape, see Fig.~\ref{FIGRandomWalks}. 
In the following, the unbiased case ($\kin=\kout\equiv k$) is of particular interest. 
In this case, there is only the single rate constant $k$, which can be 
absorbed in the unit of time, so that the distribution $P(\tc)$ depends 
only on the number of lattice points (i.e. the number of mutations). 
However, in the limit of large $M$ this dependence also disappears, if we 
use the rescaled collision time $\ttc=\tc k/M^2$. 
The resulting parameter-free distribution can be expressed in the form 
(see {\it Supporting Material}) 
\begin{equation}
  \label{critical_distribution}
  P(\ttc)=-\frac{16}{\pi^2}\,\left.\frac{\partial}{\partial t} 
  \left[Q(t)\right]^2 \right|_{t=\ttc} \;,
\end{equation}
where $Q(t)$ is the rapidly converging series 
\begin{equation}
  Q(t)=\sum_{n=1}^{\infty} \frac{(-1)^n}{2n-1}\,
       \exp\left(-\frac{\pi^2}{2}(2n-1)^2\,t\right) \;.
\end{equation}
This distribution is plotted as the solid line in the main panel of 
Fig.~\ref{FIGtwait_distribution}. Even when $M$ is small, the 
distribution (\ref{critical_distribution}) is a good approximation to 
the actual distribution, as illustrated by the dashed line in 
Fig.~\ref{FIGtwait_distribution} showing the exact distribution for 
the case of $M=7$. 

In the case of biased RW ($\kin\neq\kout$), we compute $P(\tc)$ numerically.
The dashed curves in the insets of Fig.~\ref{FIGtwait_distribution} show 
these distributions for $M=7$ mutations, where we have used the 
rates $\kin$, $\kout$ as fit parameters.

\begin{figure}[tb]
\includegraphics[width=8cm]{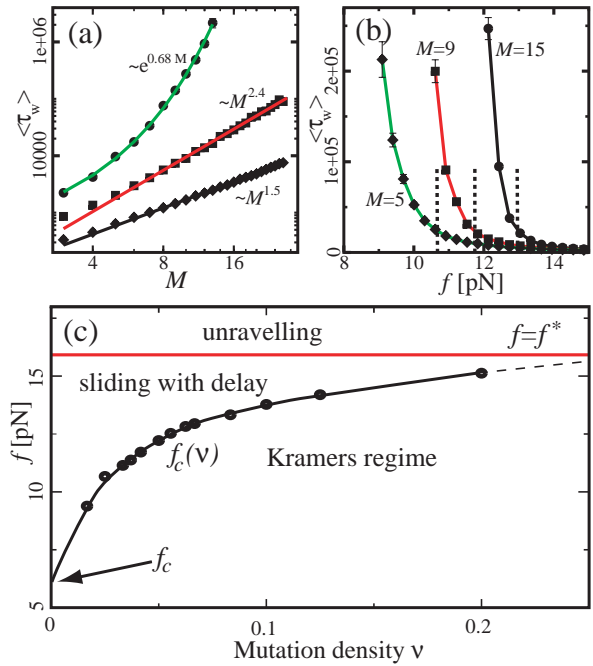}
\caption{(a) The mean waiting time $\langle\tw\rangle$ as a function of the system size (the mutation density of $\nu=1/15$ is kept fixed as the number of evenly spaced mutations $M$ is increased). At low forces the scaling is exponential (circles: data for $f=11.4$~pN; solid line: exponential fit), while we find power law behavior at the force threshold ($\tfc=12.9$~pN, squares) and above ($f=15.2$~pN, diamonds). 
(b) The mean waiting time as a function of the applied force for a sequence of $N=240$~bp with 5, 9 and 15 mutations. The dashed lines indicate the threshold force $\tfc$ for each case. Below the threshold, $\langle\tw\rangle$ rises sharply. 
(c) Different regimes of the DNA dynamics in the parameter space $(f,\nu)$. 
The Kramers regime (DNA rupture becomes exponentially slow with increasing system size) is separated from the (delayed) sliding regime by the line $\tfc(\nu)$ where the inward and outward hopping rates are equal,  $k_{in}=k_{out}$ (circles: data, solid line: interpolation). At forces larger than $\fd$, the molecule dissociates by unravelling from both ends. 
}
\label{FIGmean_twait}
\end{figure}

{\it Scaling of mean waiting time.---}
In the 2RW model, the mean first collision time follows the 
diffusive behavior $\tc\sim M^2$ when the RW are unbiased, see above. 
When the walkers have an inward bias, this changes to linear scaling 
$\tc\sim M$, whereas $\tc$ increases exponentially with $M$ for an outward 
bias, see {\it Supporting Material}. 
To test these predictions of the 2RW model, we determined 
the mean waiting time $\langle\tw\rangle$ for different $M$ and different 
forces $f$ from our DNA simulations. 
Fig.~\ref{FIGmean_twait}a shows $\langle\tw\rangle$ as a function of $M$ 
(on a double logarithmic scale) for the same three force values as in 
Fig.~\ref{FIGtwait_distribution}. 
Here, we increased the total DNA length $N$ proportional to $M$, in 
order to keep the mutation density constant and equal to that of 
Fig.~\ref{FIGtwait_distribution}. 
At the smallest force, the waiting time increases exponentially 
with $M$, as expected. 
At the intermediate force, corresponding to the case of unbiased walkers, 
we find a scaling $\langle\tw\rangle\sim M^\zeta$ with 
$\zeta\approx 2.4$, whereas $\zeta\approx 1.5$ for the largest force. 
We expect that the values of these exponents are strongly influenced by 
finite size effects, since we can vary $M$ only over roughly one decade. 
However, our results indicate that the waiting times increase more rapidly 
with the system size than expected on the grounds of our phenomenological 
2RW model. A possible explanation is given in `Microscopic mechanism' below. 

How does the mean waiting time depend on the applied force? 
Fig.~\ref{FIGmean_twait}b shows three curves of $\langle\tw\rangle$ 
vs. $f$ for different mutation densities. 
The vertical dashed lines indicate the force value where $\kin=\kout$ for 
each curve. Below these values, $\langle\tw\rangle$ increases sharply 
with decreasing force. 
Indeed, it is reasonable to consider the force $\tfc$ where $\kin=\kout$ 
as a generalization of the critical force $\fc$ to the case of weakly 
disordered sequences. 
As explained in {\it Supporting Material}, the rates $\kin$, $\kout$ 
can be extracted in several different ways from the simulation data, 
leading to $\tfc$ values which are mutually compatible. 

Fig.~\ref{FIGmean_twait}c summarizes the different dynamical regimes as 
a function of the applied force $f$ and the mutation density $\nu$. 
Without mutations ($\nu=0$) the force axis is divided into three regimes, 
with rupture driven by rare fluctuations, continuous sliding, and 
unravelling at low, intermediate, and large forces, respectively. 
As mutations are introduced ($\nu>0$), the boundary $\tfc(\nu)$ between 
the fluctuation-driven `Kramers regime' and the `sliding regime' 
rises to larger forces, and the sliding regime acquires the time delay of 
Fig.~\ref{FIGtrajectories} as a new feature. 
It is clear from Fig.~\ref{FIGmean_twait}c that the force interval 
displaying sliding behavior becomes narrower as the mutation density is 
increased. This trend can be understood within a more microscopic 
picture, see below. We could not determine unambiguously whether the 
sliding regime vanishes completely already at a finite mutation density. However, it is clear that sliding will in practice be unobservable for sequences with many mutations. 
The qualitative features depicted in Fig.~\ref{FIGmean_twait}c are 
robust against variations in our microscopic parameters $\eps$, $\Eini$, $\tbs(f)$, and $\tbd(f)$. However, the positions of the boundaries between the different regimes depend on these parameters, see below. 

{\it Microscopic mechanism.---}
Why does the mutation dynamics of Fig.~\ref{FIGtrajectories} (bottom) 
resemble the behavior of two random walkers? 
First, the opening of a mutation (and subsequent local shift of the two 
strands against each other) is always associated with the formation of 
two permanent loops, see Fig.~\ref{FIGRandomWalks}(left). 
Hence, the opening of mutations is energetically expensive and mutations 
remain mostly closed, as long as this cost is not compensated by any gain 
in entropy or stretching energy. 
Since there is no such gain when an interior mutation opens, mutations can only open beginning from the ends towards the inside: loops are constantly created at the ends of both strands and propagate inwards until they hit a mutation, which forms a barrier to bulge loop diffusion.  
On the unstretched strand, loops are generated at a higher rate than on 
the stretched strand, resulting in a larger quasi-equilibrium loop 
density \citep{Neher:04}. 
When the outermost bound mutation opens spontaneously, the accumulated loops on the exterior unstretched strand can suddenly penetrate to the inside. This penetration results in an entropy gain and a relative shift of the mutated bases, which prevents immediate recombination, see Fig.~11 in {\it Supporting Material}. 
The size of the entropy gain and the shift increases with the distance to 
the next mutation. 
Therefore, the mutation density, not the absolute number, is the 
relevant parameter that determines the relative magnitude of the hopping rates $\kin$ and $\kout$ in the random walker model, and hence fixes the value of the force threshold $\tfc$. 

We now discuss how parameter changes affect the location of the boundaries between the different dynamical regimes in Fig.~\ref{FIGmean_twait}c. First, it is clear that increasing the basepairing energy $\eps$, will shift both force thresholds, $\fc$ and $\fd$, towards higher forces. Furthermore, from the above microscopic picture, it follows that the inward hopping rate $\kin$ is proportional to the average loop density, while the outward hopping rate $\kout$ decreases with the loop density. The average loop density in turn is affected by our energy parameters: with increasing $\eps$, $\Eini$, the average loop density decreases, and consequently the boundary $\tfc(\nu)$ is shifted towards lower mutation densities, i.e. the sliding regime becomes more sensitive to mutations (this tendency is enhanced by the rising energetic cost for opening a mutation). 

So far, we considered only mutations with binding energy equal to the 
original bases. Dropping this restriction leads to a sloped boundary 
$\fd(\nu)$ between the sliding and unravelling regimes, and also 
affects the slope of $\tfc(\nu)$. 
Furthermore, we assumed above that all mutations are of a different type 
and bind only to their native binding partner. Without this assumption, 
bases belonging to different mutated basepairs can bind on encounter 
during the sliding phase. These basepairs have to be opened in the same 
way as during the waiting phase preceding sliding. 
When mutations are equidistant, this effect becomes particularly 
strong, leading to additional intervals of constant length, i.e. 
plateaus in the extension vs. time trace. 
Another important effect, caused both by variable spacing 
and energies of mutations, is that the hopping rates $\kin$ and 
$\kout$ become site dependent, so that the random walks are 
effectively on a rugged energy landscape \citep{Derrida:83,Kafri:04}. 

Finally, we stress that the 2RW model is phenomenological 
and fails to describe certain features of the DNA dynamics. 
For instance, our simple description has neglected correlations between 
subsequent hopping steps of a RW, see {\it Supporting Material}. 
Short-range correlations do not affect the long-time behavior, which 
may explain why our model describes the shape of the waiting time 
distribution accurately, see Fig.~\ref{FIGtwait_distribution}. 
A more drastic approximation is that the 2RW model does not 
account for the time required to bring in new loops from the ends 
to a mutation deep inside the dsDNA. 
The fact that this time increases with the length of the DNA may be the 
cause for the waiting time to rise more rapidly with the system size than 
expected from the 2RW model, see Fig.~\ref{FIGmean_twait}.

\section*{Conclusions and Outlook} 

The basepairing dynamics in DNA and RNA molecules is only beginning to be 
explored. Here, we have shown that even the seemingly simple case of 
periodic DNA sequences displays rich behavior, which can be revealed by 
applying a shear force. Our main finding is that the microscopic dynamics of bulge loop defects endows DNA with viscoelastic properties, which can be programmed into the sequence. Weak sequence disorder does not abolish these properties, but (i) introduces a delay, since all mutations have to be broken before DNA sliding begins, and (ii) effectively narrows the viscoelastic force regime. The dynamics of mutation breaking is an interesting statistical process, with main features that can be understood by considering a first passage problem of two random walkers. 
Our theoretical study has led to several experimental ramifications. For instance, we predict that periodic or nearly periodic DNA responds to sudden stress by slowly relaxing its tension to a threshold value independent of the initial stress (provided the DNA is not too short). This stress relaxation process cannot occur for heterogeneous DNA. 
Furthermore, we predict that the relaxation velocity is inversely proportional to the DNA length, so that the timescale of the dynamics can 
be easily adjusted into the range of interest for a given experimental setup. We expect the existence of the different dynamical regimes shown in Fig.~\ref{FIGmean_twait}c to be independent of our detailed model assumptions. As DNA slippage is directly linked to the production rate and mobility of bulge loops, single molecule experiments on DNA sliding would test our basic understanding of basepairing dynamics in DNA. 

The same properties, that make DNA uniquely suited for reliably storing 
genetic information while keeping it accessible, permit many applications 
in nanotechnology \citep{Seeman:03}.  
For instance, dsDNA has been used as reversible crosslinker in polymer networks to switch between different mechanical properties \citep{Lin:04}, 
and even DNA-only networks with specified topologies can be constructed,  exploiting the specificity of the basepairing interaction \citep{Seeman:03}.
In other applications, short dsDNA molecules served as programmable force sensors \citep{Albrecht:03} using the sequence-dependence of the mechanical rupture force, or DNA-based nano-machines were constructed on the basis of the DNA branch migration mechanism \citep{Simmel:02}.
Our results render several new applications for DNA in nanotechnology conceivable. For instance, complementary periodic ssDNA's could be used as self-tightening connections in nanostructures: once two such strands found each other, they will slide to maximize their overlap until the tension reaches a value $\fc$. Periodic or nearly periodic DNA could also serve as a viscoelastic crosslinker in polymer networks, which should lead to different material properties from those observed in \citep{Lin:04}. Similarly, DNA networks could also be endowed with viscoelastic properties, and (nearly) periodic DNA might even be useful as a programmable reference molecule for kinetic measurements. Of course, which of these and other possible applications will turn out to be useful in the end is unclear at the present stage. However, we feel that there is a clear potential that should be explored.

{\it Acknowledgments.---}
We thank T.~Hwa, K.~Kroy, F.~K\"uhner, and F.~Simmel for useful discussions and the {\it DFG} for financial support.

\end{document}